# Some new parallel flows due to Lorentz forces in electrically conducting fluids

Asterios Pantokratoras
Associate Professor of Fluid Mechanics
School of Engineering, Democritus University of Thrace,
67100 Xanthi – Greece

e-mail:apantokr@civil.duth.gr

Abstract

We investigate the fully developed flow between two parallel plates and the film flow over a plate in an electrically conducting fluid under the action of a parallel Lorentz force. Exact analytical solutions are derived for velocity, flow rate and wall shear stress at the plates. The velocity results are presented in figures. All these flows are new and are presented for the first time in the literature.

## INTRODUCTION

In 1961 Gailitis and Lielausis introduced the idea of using a Lorentz force to control the flow of an electrically conducting fluid over a flat plate. This is achieved by applying an external electromagnetic field (see figure 1) by a stripwise arrangement of flush mounted electrodes and permanent magnets of alternating polarity and magnetization. The Lorentz force, which acts parallel to the plate, can either assist or oppose the flow. This idea was later abandoned and only recently attracted new attention (Henoch and Stace, 1995, Crawford and Karniadakis, 1997, Berger et al. 2000). In addition, in last years much investigation on flow control using the Lorentz force is being conducted at the Rossendorf Institute and at the Institute for Aerospace Engineering in Dresden, Germany (Posdziech and Grundmann, 2001, Weier and Gerbeth, 2004, Weier, 2005, Mutschke et al. 2006, Albrecht and Grundmann, 2006).

Although the Lorentz force has been applied in many flow configurations, some simple flows under the influence of this force have not been investigated until now. In the present paper we will investigate the classical flow between two parallel, infinite plates and the problem of a film flowing over a plate.

## THE MATHEMATICAL MODEL

Consider the flow between two horizontal, infinite, parallel plates with $u$ and $v$ denoting respectively the velocity components in the $x$ and $y$ direction, where $x$ is the coordinate along the plates and $y$ is the coordinate perpendicular to $x$. It is assumed that an electromagnetic field exists at the lower plate and therefore a Lorentz force, parallel to the plates, is produced. The fluid is forced to move due to the action of the Lorentz force. For steady, two-dimensional flow the boundary layer equations with constant fluid properties are (Tsinober and Shtern, 1967, Weier, 2005)

continuity equation: $$\frac{\partial u}{\partial x} + \frac{\partial v}{\partial y} = 0 \qquad (1)$$

momentum equation: $$u\frac{\partial u}{\partial x} + v\frac{\partial u}{\partial y} = -\frac{1}{\rho}\frac{\partial p}{\partial x} + \nu\frac{\partial^2 u}{\partial y^2} + \frac{\pi j_0 M_0}{8\rho}\exp(-\frac{\pi}{a}y) \qquad (2)$$

where $p$ is the pressure, $v$ is the fluid kinematic viscosity, $j_0$ (A/m$^2$) is the applied current density in the electrodes, $M_0$ (Tesla) is the magnetization of the permanent magnets, $a$ is the width of magnets and electrodes and $\rho$ is the fluid density. The last term in the momentum equation is the Lorentz force which decreases exponentially with $y$ and is independent of the flow. For fully developed conditions the flow is parallel, the transverse velocity is zero, and the flow is described only by the following momentum equation

momentum equation: $$-\frac{1}{\rho}\frac{dp}{dx}+v\frac{\partial^2 u}{\partial y^2}+\frac{\pi j_0 M_0}{8\rho}\exp(-\frac{\pi}{a}y)=0 \qquad (3)$$

RESULTS AND DISCUSSION

1. **The classical Couette flow with Lorentz force**

The first viscous fluid flow treated in the classical book by White (2006) is the steady flow between a fixed and a moving plate (Couette flow) and this happens in almost all Fluid Mechanics books because this flow is the simplest in Fluid Mechanics (Liggett, 1994, page 156, Kleinstreuer, 1997, page 121, Panton, 2005, page 132). This flow is called Couette flow in honour of the French M. Couette (1890) who performed experiments on the flow between a fixed and moving concentric cylinder. The boundary conditions for this case are:

at $y = 0$: $u = 0$ (4)
as $y=h$: $u = u_2$ (5)

where $h$ is the distance between the plates and $u_2$ is the velocity of the upper plate.

Here we will investigate this flow under the action of a Lorentz force produced at the lower plate. The momentum equation (3), without the pressure gradient, with boundary conditions (4)-(5) has the following exact analytical solution

$$\frac{u}{u_2}=\frac{y}{h}+Z\left[1-\exp(-\frac{\pi}{a}y)-\frac{y}{h}(1-\exp(-\frac{\pi}{a}h))\right] \qquad (6)$$

The first term on the right is due to the motion of the upper plate (classical

Couette flow) and the second term due to the action of the Lorentz force. The parameter $Z$ is the Lorentz number defined as

$$Z = \frac{j_0 M_0 a^2}{8\pi\mu u_2} \tag{7}$$

where $\mu$ is the fluid dynamic viscosity. The Lorentz number is dimensionless and expresses the balance between the electromagnetic forces to viscous forces. This number is used in the analysis of the boundary layer flow over a flat plate situated in a free stream and there as characteristic velocity is used the free stream velocity (Weier, 2005). We see that this number appears also in the Couette flow with characteristic velocity that of the moving plate. The dimensionless velocity given by equation (6) depends both on Lorentz number $Z$ and the ratio $h/a$. The above combination of the classical Couette flow with Lorentz forces is presented here for the first time in the literature and may be called Couette-Pantokratoras flow.

The dimensionless flow rate $M$ between the plates is

$$M = \frac{1}{u_2 h}\int_0^h u\,dy \tag{8}$$

and is obtained by integrating the velocity function. Thus we have

$$M = \frac{1}{2} + \frac{Z}{2\pi}\left[(\pi + 2\frac{a}{h})\exp(-\frac{\pi}{a}h) + \pi - 2\frac{a}{h}\right] \tag{9}$$

The wall shear stress at the two plates are

$$\tau_1 = \frac{\mu u_2}{h} + \mu u_2 Z\left[\frac{\pi}{a} - \frac{1}{h}(1 - \exp(-\frac{\pi}{a}h))\right] \tag{10}$$

$$\tau_2 = \frac{\mu u_2}{h} + \mu u_2 Z\left[-\frac{1}{h} + (\frac{\pi}{a} + \frac{1}{h})\exp(-\frac{\pi}{a}h)\right] \tag{11}$$

The wall shear stress $\tau_1$ becomes zero when the Lorentz number takes the value

$$Z=\left[-\frac{\pi h}{a}+1-\exp(-\frac{\pi}{a}h)\right]^{-1} \tag{12}$$

while $\tau_2$ becomes zero when

$$Z=\left[1-(\frac{\pi h}{a}+1)\exp(-\frac{\pi}{a}h)\right]^{-1} \tag{13}$$

In figure 2 some velocity profiles are presented for different values of the Lorentz number and *h/a*=1. The profile 1 corresponds to zero shear stress at the lower plate and this happens when Z=-0.4577064. The profile 3 corresponds to zero shear stress at the upper plate and this happens when Z=1.2179889. The profile 2 corresponds to Couette flow.

When the lower plate, where the electromagnetic field is produced, is moving and the upper plate is motionless the velocity is given by the following equation

$$\frac{u}{u_1}=1-\frac{y}{h}+Z\left[1-\exp(-\frac{\pi}{a}y)-\frac{y}{h}(1-\exp(-\frac{\pi}{a}h))\right] \tag{14}$$

where $u_1$ is the velocity of the lower plate and the Lorentz number is based on velocity of the lower plate. Now the flow rate is defined as

$$M=\frac{1}{u_1 h}\int_0^h u\,dy \tag{15}$$

and equation (9) is valid also for this case. The wall shear stress at the two plates are

$$\tau_1=-\frac{\mu u_1}{h}+\mu u_1 Z\left[\frac{\pi}{a}-\frac{1}{h}(1-\exp(-\frac{\pi}{a}h))\right] \tag{16}$$

$$\tau_2=-\frac{\mu u_1}{h}+\mu u_1 Z\left[-\frac{1}{h}+(\frac{\pi}{a}+\frac{1}{h})\exp(-\frac{\pi}{a}h)\right] \tag{17}$$

The wall shear stress $\tau_1$ becomes zero when the Lorentz number takes the value

$$Z = \left[ \frac{\pi h}{a} - 1 + \exp(-\frac{\pi}{a} h) \right]^{-1} \tag{18}$$

while $\tau_2$ becomes zero when

$$Z = \left[ -1 + (\frac{\pi h}{a} + 1)\exp(-\frac{\pi}{a} h) \right]^{-1} \tag{19}$$

In figure 3 some velocity profiles are presented for different values of the Lorentz number and h/a=1. The profile 1 corresponds to zero shear stress at the upper plate and this happens when Z=-1.2179889. The profile 3 corresponds to zero shear stress at the lower plate and this happens when Z=0.4577064. The profile 2 corresponds to Couette flow.

## 2. **The classical Poiseuille flow with Lorentz forces**

Another kind of flow between parallel plates is the Poiseuille flow (Poiseuille, 1840) which is caused by a constant pressure gradient along the plates while the plates are motionless. This flow is also included in Fluid Mechanics books (Liggett, 1994, page 157, Panton, 2005, page 125, White, 2006, page 106). The boundary conditions are

at $y = 0$: $u = 0$ (20)

as $y=h$: $u=0$ (21)

The analytical solution of equation (3) with boundary conditions (20)-(21) is

$$u = -\frac{h^2}{2\mu}\frac{dp}{dx}\frac{y}{h}(1-\frac{y}{h}) + \frac{j_0 M_0 a^2}{8\pi\mu}\left[1 - \exp(-\frac{\pi}{a}y) - \frac{y}{h}(1-\exp(-\frac{\pi}{a}h))\right] \tag{22}$$

The first term on the right is due to the pressure gradient (classical Poiseuille flow) and the second term due to the action of the Lorentz force. In the above equation there are two characteristic velocities, the first one due to pressure gradient and the second due to Lorentz forces as follows

$$u_P = -\frac{h^2}{2\mu}\frac{dp}{dx} \tag{23}$$

$$u_Z = \frac{j_0 M_0 a^2}{8\pi\mu} \tag{24}$$

Taking into account these characteristic velocities equation (22) becomes

$$u = u_P \frac{y}{h}(1-\frac{y}{h}) + u_Z\left[1-\exp(-\frac{\pi}{a}y) - \frac{y}{h}(1-\exp(-\frac{\pi}{a}h))\right] \tag{25}$$

and in dimensionless form

$$\frac{u}{u_Z} = \frac{u_P}{u_Z}\frac{y}{h}(1-\frac{y}{h}) + \left[1-\exp(-\frac{\pi}{a}y) - \frac{y}{h}(1-\exp(-\frac{\pi}{a}h))\right] \tag{26}$$

The ratio $u_P/u_Z$ is a new dimensionless number which expresses the balance between the pressure forces to electromagnetic forces

$$Pa_p = \frac{u_P}{u_Z} \tag{27}$$

The dimensionless velocity given by equation (26) is a function of $Pa_p$ and $h/a$. The above combination of the classical Poiseuille flow with Lorentz forces is a new kind of parallel flow and may be called Poiseuille-Pantokratoras flow. We define the dimensionless flow rate as

$$M = \frac{1}{u_Z h}\int_0^h u\,dy \tag{28}$$

and M is

$$M = \frac{Pa_p}{6} + \frac{1}{2\pi}\left[(\pi + 2\frac{a}{h})\exp(-\frac{\pi}{a}h) + \pi - 2\frac{a}{h}\right] \tag{29}$$

The wall shear stress at the two plates are

$$\tau_1 = \frac{\mu u_P}{h} + \mu u_Z\left[\frac{\pi}{a} - \frac{1}{h}(1-\exp(-\frac{\pi}{a}h))\right] \tag{30}$$

$$\tau_2 = -\frac{\mu u_P}{h} + \mu u_Z\left[-\frac{1}{h} + (\frac{\pi}{a} + \frac{1}{h})\exp(-\frac{\pi}{a}h)\right] \tag{31}$$

The wall shear stress $\tau_1$ becomes zero when the quantity $Pa_p$ takes the value

$$Pa_p = \left[-\frac{\pi h}{a} + 1 - \exp(-\frac{\pi}{a}h)\right] \tag{32}$$

while $\tau_2$ becomes zero when

$$Pa_p = \left[-1 + (\frac{\pi h}{a} + 1)\exp(-\frac{\pi}{a}h)\right] \tag{33}$$

In figure 4 we present some velocity profiles for $h/a$=1 and different $Pa_p$ values. The profile 1 corresponds to zero shear stress at the lower plate and this happens when $Pa_p$=-2.1848066. The profile 3 corresponds to zero shear stress at the upper plate and this happens when $Pa_p$=-0.8210255. The profile 4 corresponds to zero pressure gradient. The profiles 1, 2 and 3 are S-shaped and each of them has an inflection point.

## 3. Flow between parallel plates due to Lorentz force only

If both plates are motionless and the pressure gradient is zero we have a flow caused by the Lorentz force only. From equation (22) we get the velocity of this flow by putting the pressure gradient zero. Thus we have

$$u = \frac{j_0 M_0 a^2}{8\pi\mu}\left[1 - \exp(-\frac{\pi}{a}y) - \frac{y}{h}(1-\exp(-\frac{\pi}{a}h))\right] \tag{34}$$

and in dimensionless form

$$\frac{u}{u_Z} = \left[1 - \exp(-\frac{\pi}{a}y) - \frac{y}{h}(1-\exp(-\frac{\pi}{a}h))\right] \tag{35}$$

This flow is completely new, it is presented here for the first time in the literature and may be called Pantokratoras flow. The dimensionless flow rate is defined as

$$M = \frac{1}{u_Z h} \int_0^h u dy \tag{36}$$

and the flow rate is

$$M = \frac{1}{2\pi} \left[ (\pi + 2\frac{a}{h}) \exp(-\frac{\pi}{a} h) + \pi - 2\frac{a}{h} \right] \tag{37}$$

The wall shear stress at the two plates are

$$\tau_1 = \mu u_Z \left[ \frac{\pi}{a} - \frac{1}{h} (1 - \exp(-\frac{\pi}{a} h) \right] \tag{38}$$

$$\tau_2 = \mu u_Z \left[ -\frac{1}{h} + (\frac{\pi}{a} + \frac{1}{h}) \exp(-\frac{\pi}{a} h) \right] \tag{39}$$

In figure 5 velocity profiles are shown for different values of *h/a*. This flow has some special characteristics. When *h/a* →∞ the maximum dimensionless velocity tends to 1 and the velocity profile tends to compose from two straight lines, one of them horizontal and the other with inclination equal to 45 degrees. When *h/a* →0 the dimensionless velocity tends to 0 and the velocity profile tends to become symmetric with its maximum at the centerline between the plates. We see also that the velocity maximum moves to the centerline as *h/a* decreases.

## 4. **Film flow due to Lorentz forces**

Another kind of simple parallel flow is that of a film falling down an inclined wall due to gravity and due to the action of a constant shear stress on the free surface (Bird et al. 2002, page 45, Panton, 2005, page 135). Here we will treat the motion of a film due to Lorentz force ignoring gravity and retaining the action of the surface shear stress. The boundary conditions for this case are:

$$\text{at } y = 0: \quad u = 0 \tag{40}$$

as $y=h$: $\mu \frac{\partial u}{\partial y} = \tau_2$ (41)

where $\tau_2$ is known and constant. The analytical solution of equation (3), without pressure gradient, with boundary conditions (40)-(41) is

$$u = \frac{j_0 M_0 a^2}{8\pi\mu}\left[1 - \exp(-\frac{\pi}{a}y)\right] + \left[\frac{\tau_2}{\mu} - \frac{j_0 M_0 a}{8\mu}\exp(-\frac{\pi}{a}h)\right]y \tag{42}$$

and in dimensionless form

$$\frac{u}{u_Z} = \left[1 - \exp(-\frac{\pi}{a}y)\right] + \left[\frac{\tau_2 h}{\mu u_Z} - \frac{\pi h}{a}\exp(-\frac{\pi}{a}h)\right]\frac{y}{h} \tag{43}$$

The new dimensionless number is

$$Pa_f = \frac{\tau_2 h}{\mu u_Z} \tag{44}$$

and equation (43) becomes

$$\frac{u}{u_Z} = \left[1 - \exp(-\frac{\pi}{a}y)\right] + \left[Pa_f - \frac{\pi h}{a}\exp(-\frac{\pi}{a}h)\right]\frac{y}{h} \tag{45}$$

The dimensionless flow rate is defined as

$$M = \frac{1}{u_Z h}\int_0^h u\,dy \tag{46}$$

and the flow rate is

$$M = \frac{a}{\pi h}\left[\exp(-\frac{\pi}{a}h) - 1\right] + \frac{1}{2}\left[Pa_f - \frac{\pi h}{a}\exp(-\frac{\pi}{a}h)\right] + 1 \tag{47}$$

The shear stress at the plate is

$$\tau_1 = \frac{j_0 M_0 a}{8}\left[1 - \exp(-\frac{\pi}{a}h)\right] + \tau_2 \tag{48}$$

The wall shear stress $\tau_1$ becomes zero when the quantity $Pa_f$ takes the value

$$Pa_f = \frac{\pi h}{a}\left[\exp(-\frac{\pi}{a}h) - 1\right] \tag{49}$$

In figure 6 velocity profiles are shown for different values of $Pa_f$ number and $h/a$=1. The curve 1 corresponds to zero wall shear stress and this happens when $Pa_f$=-3.0058321 while curve 2 corresponds to zero surface shear stress ($Pa_f$=0), that is, the flow is produced by the Lorentz force only.

5. **Film flow due to Lorentz force only**

If the shear stress at the surface is zero we have a film flow caused by the Lorentz force only. From equation (45) we get the velocity of this flow by putting $Pa_f$=0. Thus we have

$$\frac{u}{u_Z} = \left[1 - \exp(-\frac{\pi}{a}y)\right] - \left[\frac{\pi h}{a}\exp(-\frac{\pi}{a}h)\right]\frac{y}{h} \tag{50}$$

The dimensionless flow rate is defined as

$$M = \frac{1}{u_Z h}\int_0^h u dy \tag{51}$$

and the flow rate is

$$M = \frac{a}{\pi h}\left[\exp(-\frac{\pi}{a}h) - 1\right] - \frac{1}{2}\left[\frac{\pi h}{a}\exp(-\frac{\pi}{a}h)\right] + 1 \tag{52}$$

The shear stress at the plate is

$$\tau_1 = \frac{j_0 M_0 a}{8}\left[1 - \exp(-\frac{\pi}{a}h)\right] \tag{53}$$

In figure 7 velocity profiles are shown for different values of $h/a$. All velocity profiles have zero gradient at the surface and meet the surface vertically. When $h/a \to\infty$ the maximum dimensionless velocity, which lies on the free surface, tends to 1 and the velocity profile tends to compose from two straight lines, one of them horizontal and the other vertical. When $h/a \rightarrow 0$ the dimensionless velocity tends to 0.

## CONCLUSIONS

In this paper some new kinds of parallel flows have been presented and analyzed for electrically conducting fluids. Exact analytical solutions have been given for velocity, flow rate and wall shear stresses. The author believes that the results of the present work will enrich the list with the existing exact solutions of the Navier-Stokes equations and may help the investigation of flow of electrically conducting fluids like water and liquid metals.

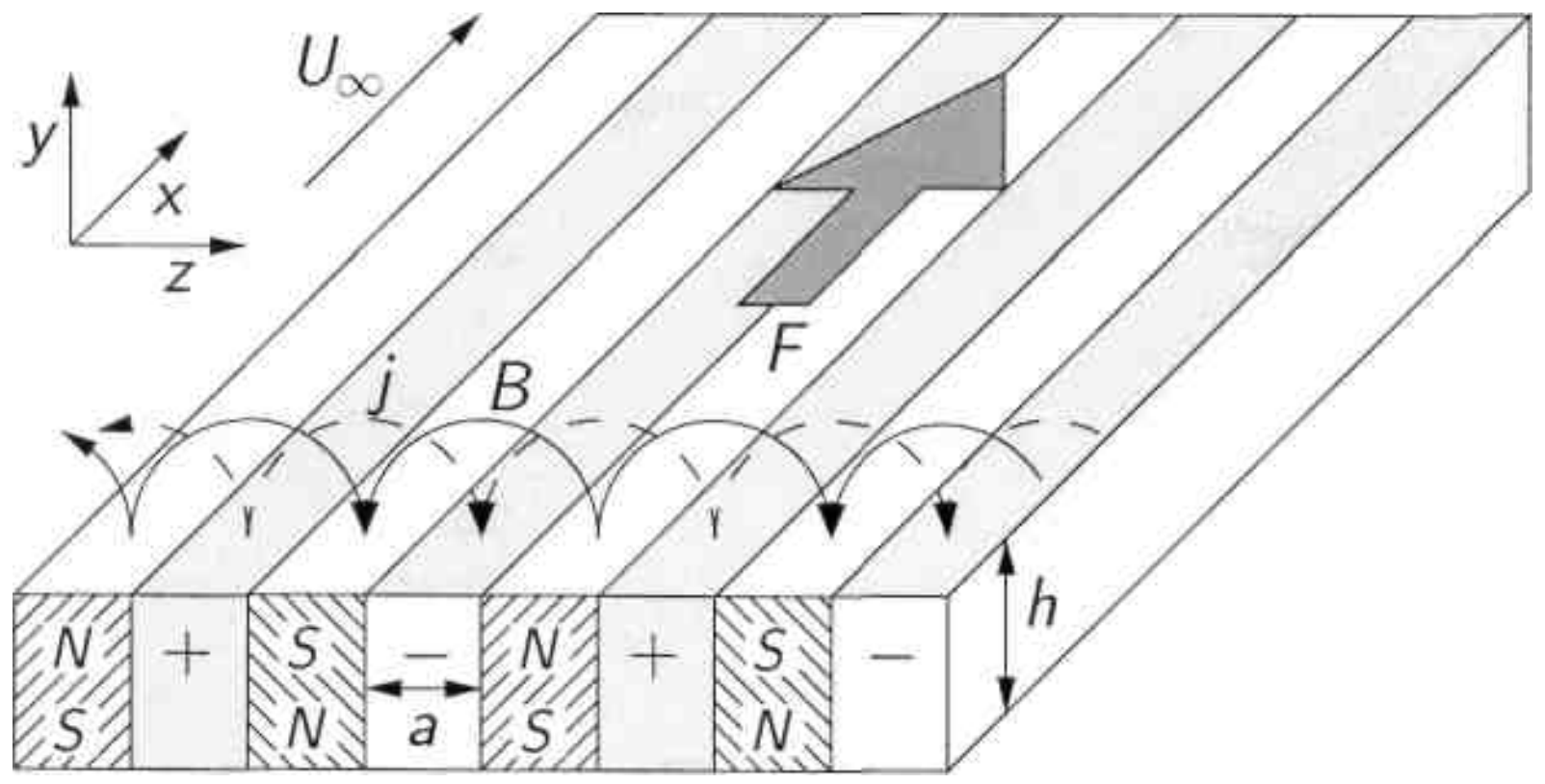

Figure 1. Arrangement of electrodes and magnets for the creation of a Lorentz force F in the flow along a flat plate (Weier, 2005).

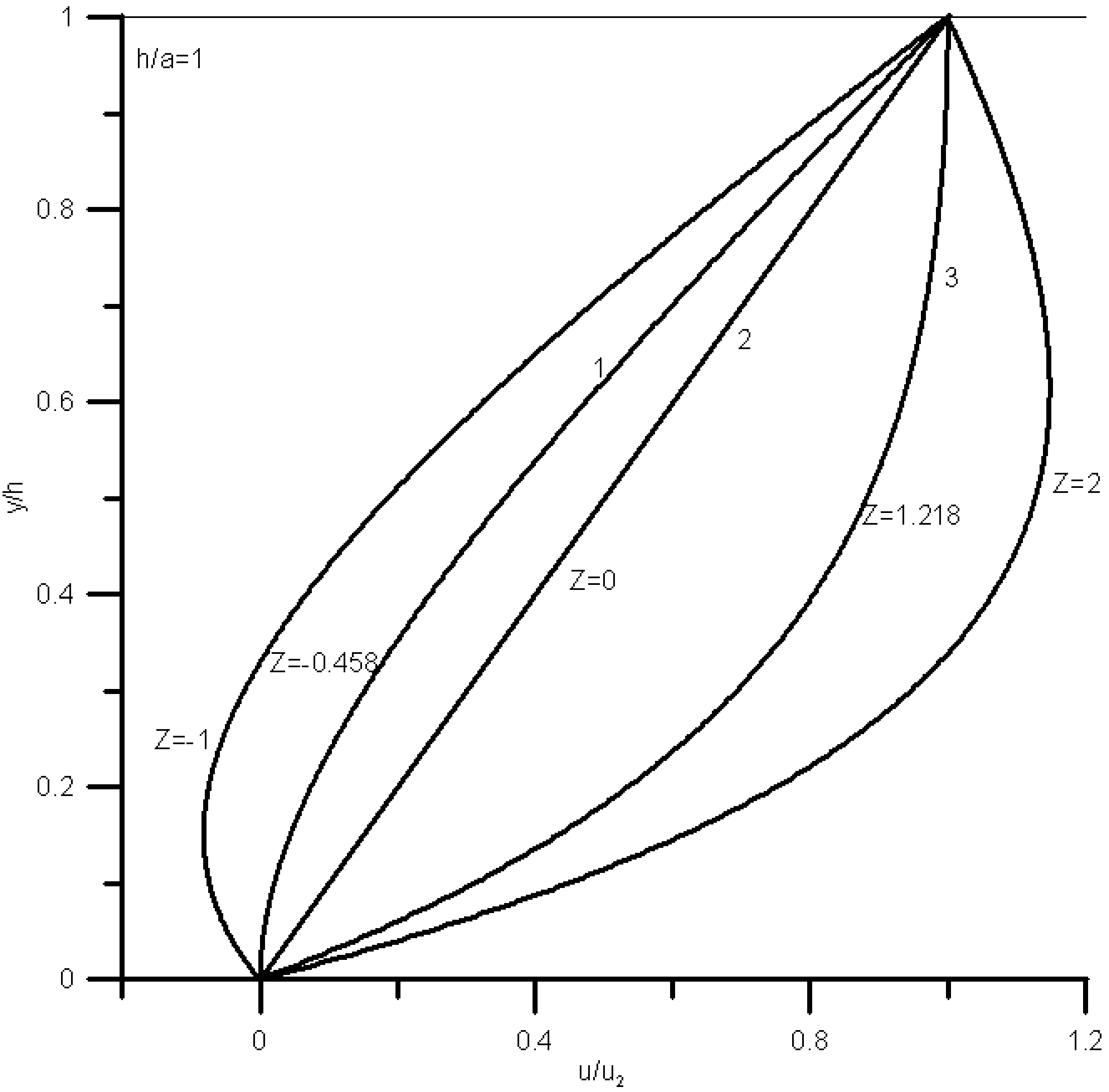

Figure 2. Velocity profiles for Couette-Pantokratoras flow for h/a=1 and different values of the Lorentz number. The case Z=0 corresponds to Couette flow with upper plate moving.

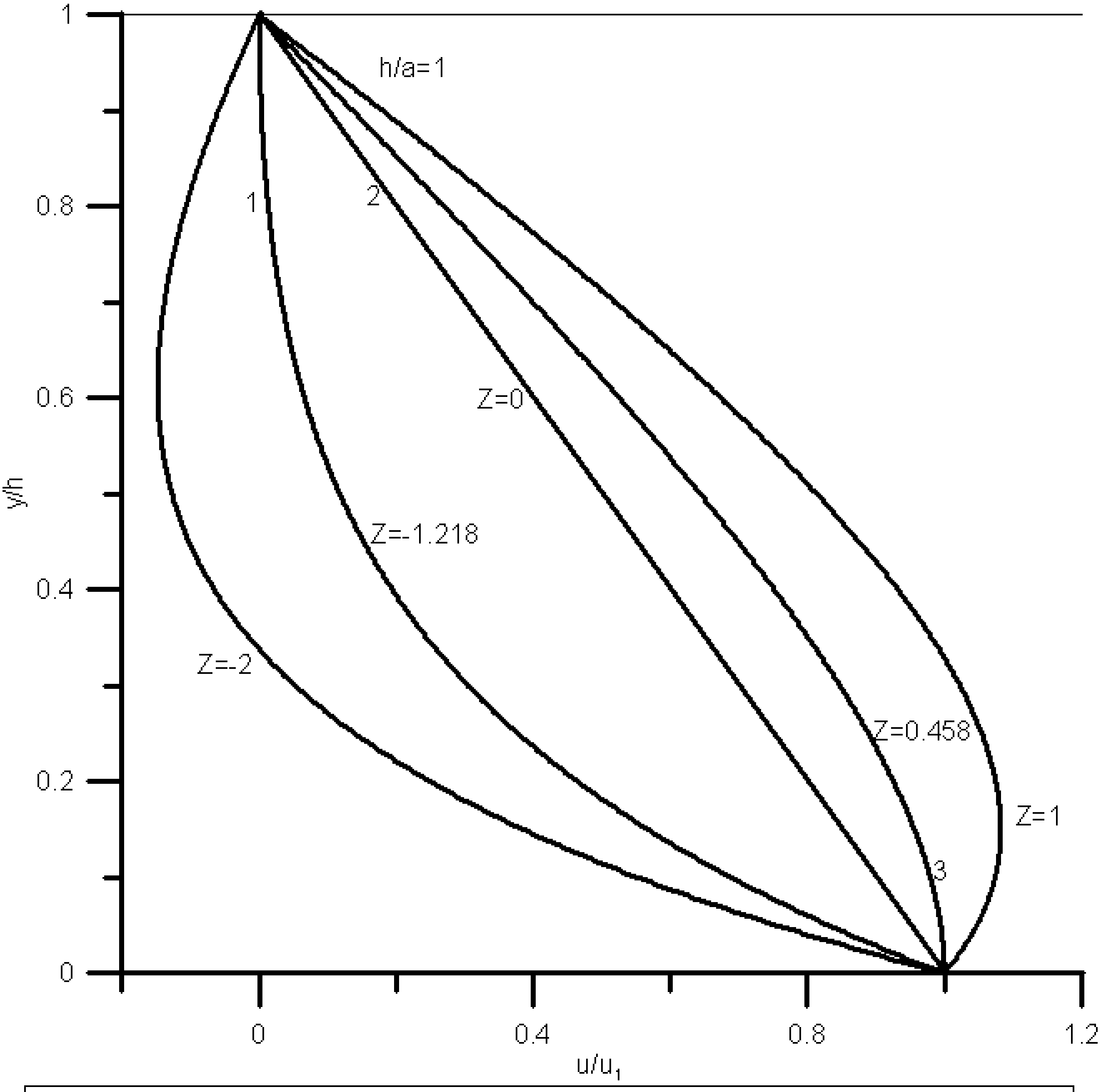

Figure 3. Velocity profiles for Couette-Pantokratoras flow for h/a=1 and different values of the Lorentz number. The case Z=0 corresponds to Couette flow with lower plate moving.

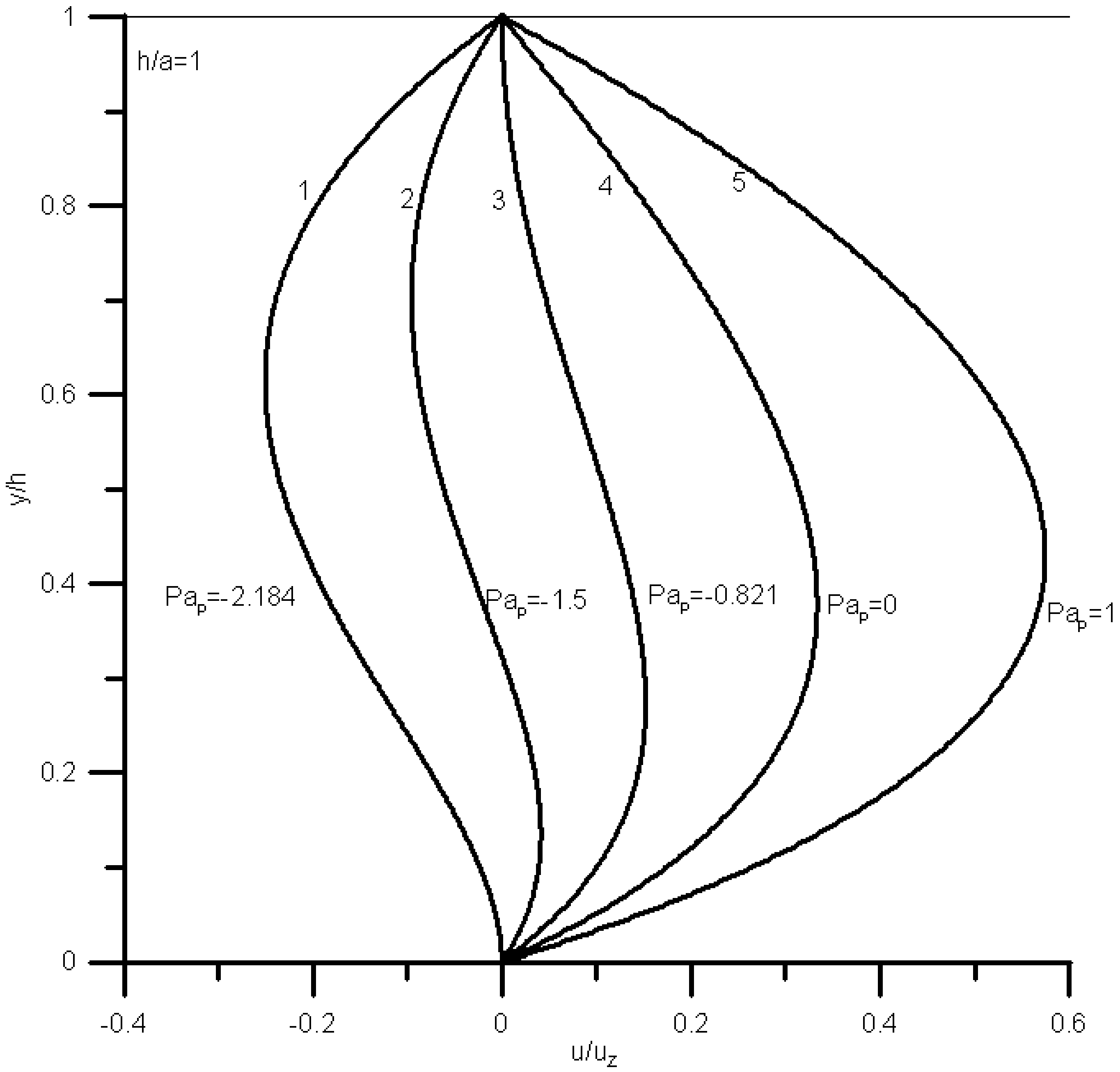

Figure 4. Velocity profiles for Poiseuille-Pantokratoras flow for h/a=1 and different values of the $Pa_p$ parameter. The case $Pa_p$=0 corresponds to zero pressure gradient.

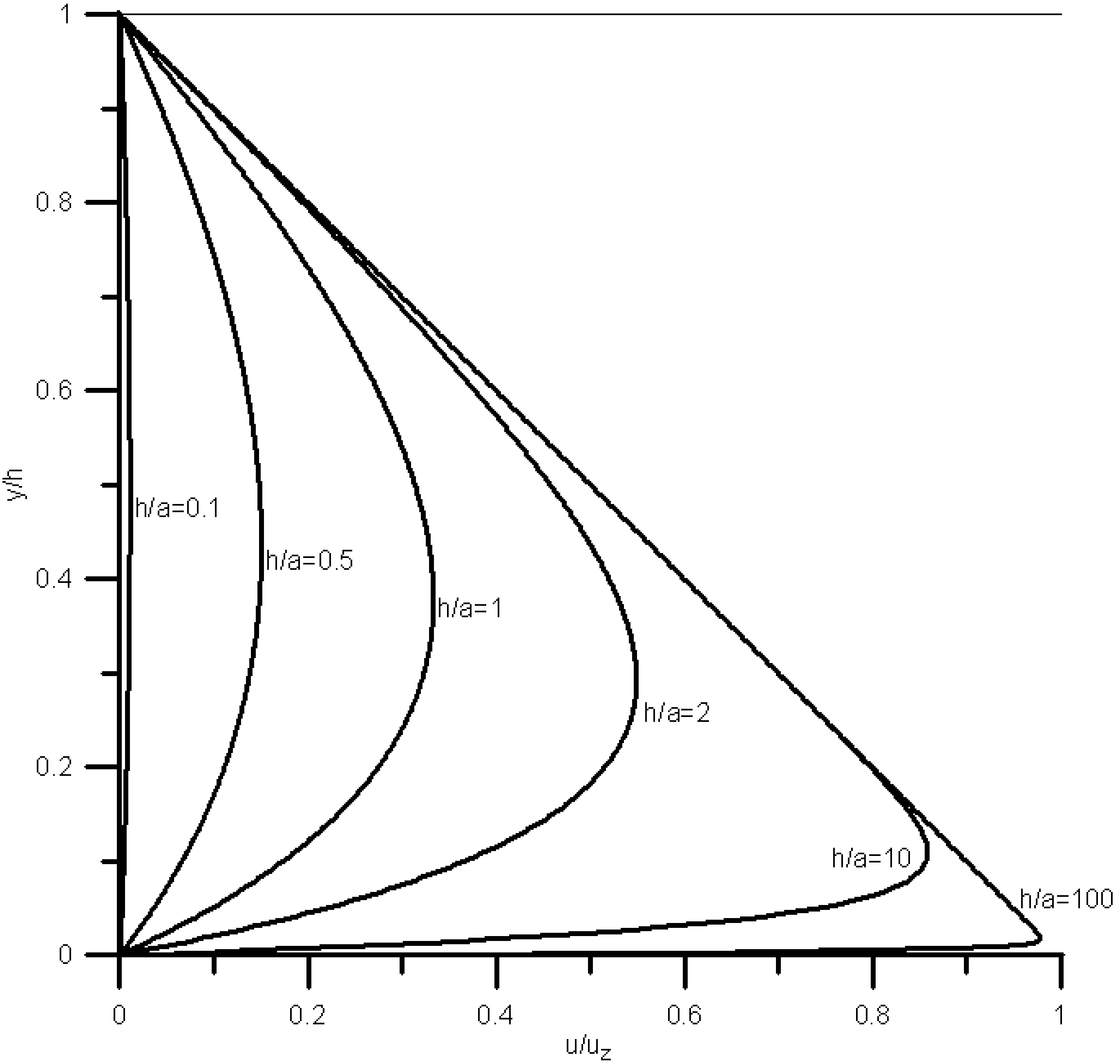

Figure 5. Velocity profiles for flow due to Lorentz force only (Pantokratoras flow) for different values of h/a.

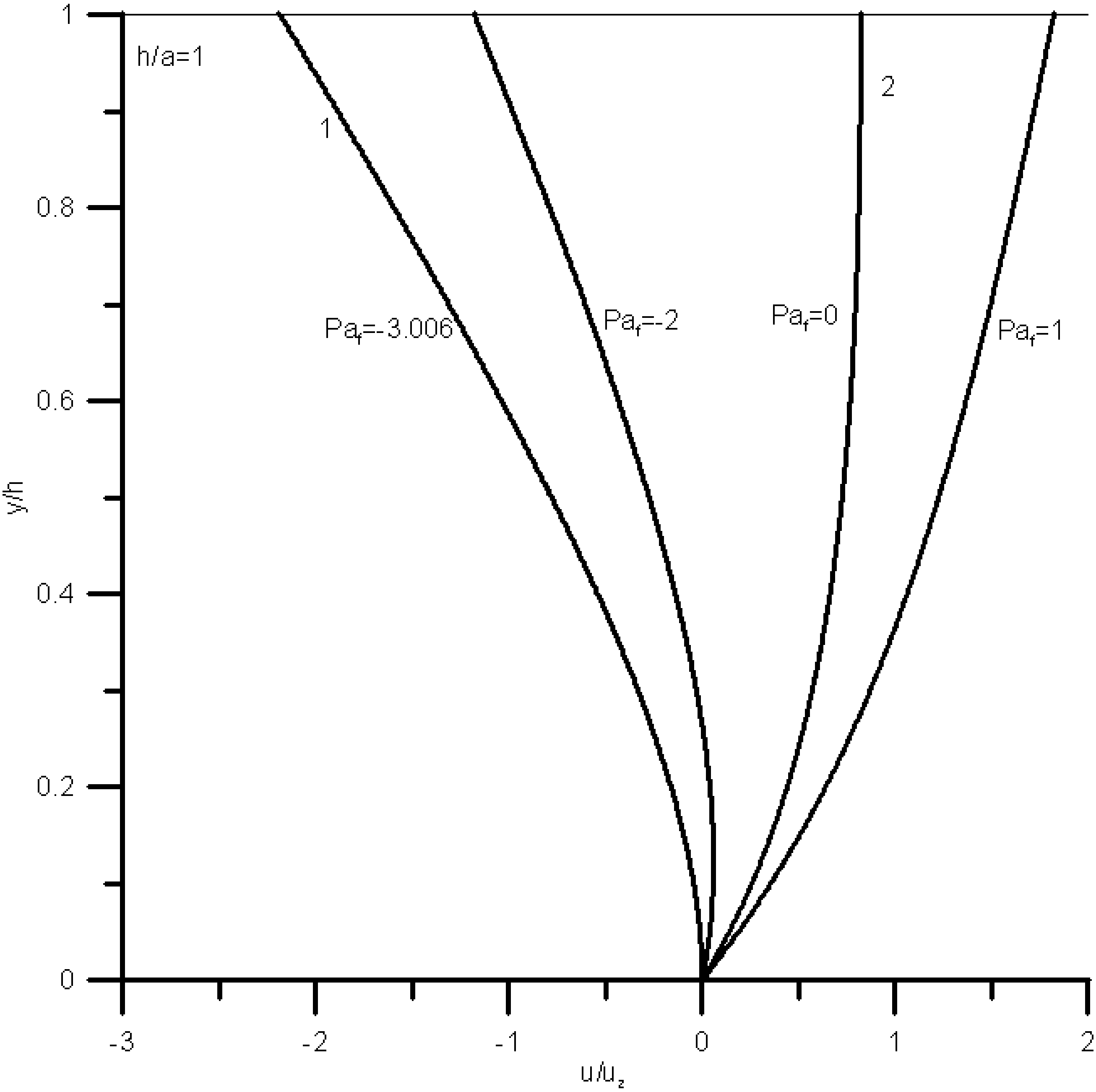

Figure 6. Velocity profiles for film flow for h/a=1 and different values of the $Pa_f$ number. The case $Pa_f=0$ corresponds to zero surface shear stress.

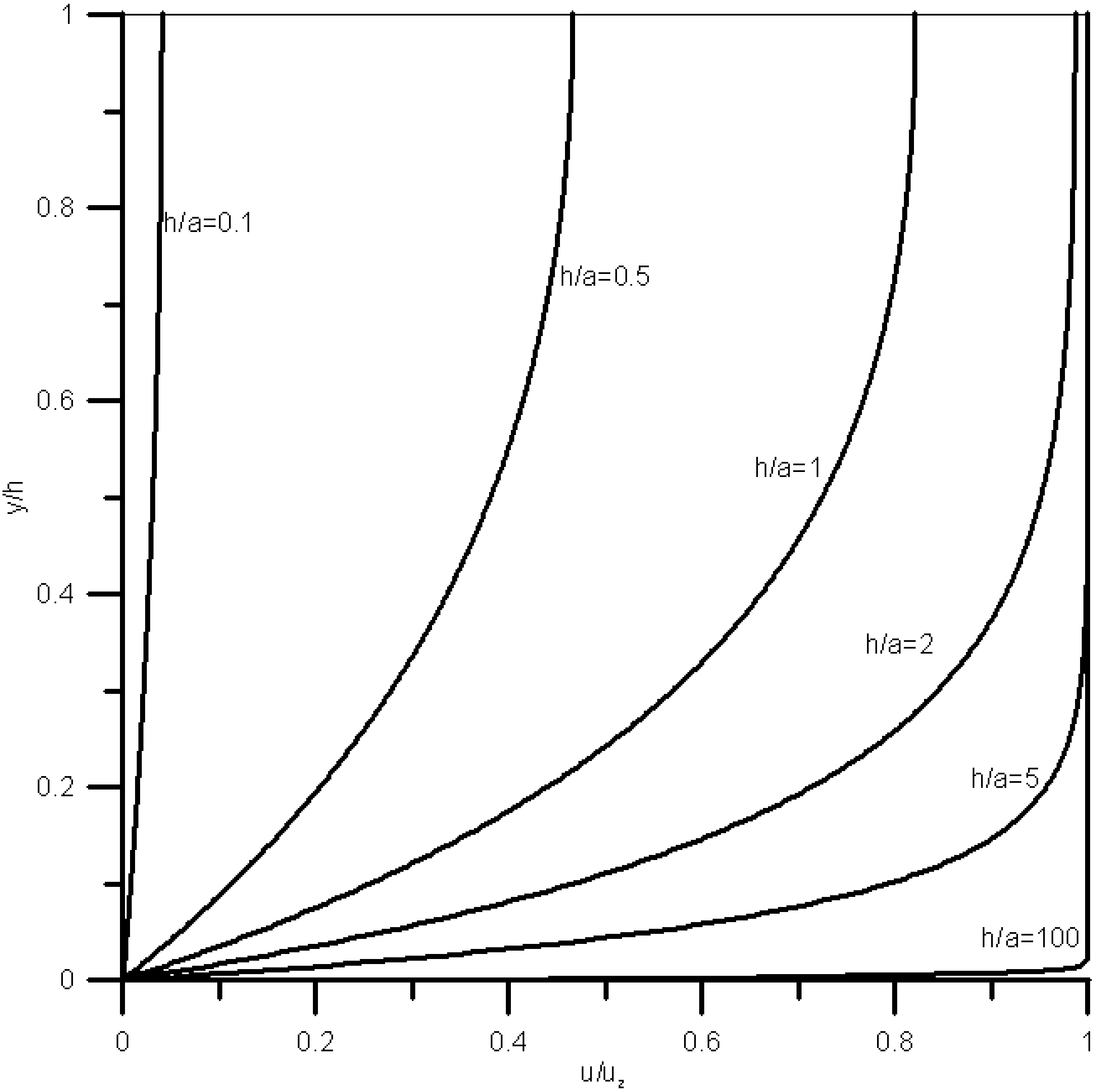

Figure 7. Velocity profiles for film flow due to Lorentz force only for different values of h/a.